\documentclass[a4paper,fleqn,usenatbib]{mnras}

\usepackage{newtxtext,newtxmath}

\usepackage[T1]{fontenc}
\usepackage{ae,aecompl}

\usepackage{graphicx}   
\usepackage{amsmath}    
\usepackage{amssymb}    

\title[Two-step solar filament eruptions]{Two-step solar filament eruptions}

\author[B. Filippov]{B. Filippov \thanks{E-mail:
bfilip@izmiran.ru}  \\ Pushkov Institute of Terrestrial Magnetism,
Ionosphere and Radio Wave Propagation of the Russian Academy of
Sciences (IZMIRAN), \\ Troitsk, Moscow 108840, Russia}

\date{Accepted XXX. Received YYY; in original form ZZZ}

\pubyear{2017}

\begin{document}
\label{firstpage}
\pagerange{\pageref{firstpage}--\pageref{lastpage}}
\maketitle

\begin{abstract}
Coronal mass ejections (CMEs) are closely related to eruptive
filaments and usually are the continuation of the same eruptive
process into the upper corona. There are failed filament eruptions
when a filament decelerates and stops at some greater height in
the corona. Sometimes the filament after several hours starts to
rise again and develops into the successful eruption with a CME
formation. We propose a simple model for the interpretation of
such two-step eruptions in terms of equilibrium of a flux rope in
a two-scale ambient magnetic field. The eruption is caused by a
slow decrease of the holding magnetic field. The presence of two
critical heights for the initiation of the flux-rope vertical
instability allows the flux rope to stay after the first jump some
time in a metastable equilibrium near the second critical height.
If the decrease of the ambient field continues, the next eruption
step follows.
\end{abstract}




\section{Introduction}

Solar filaments, or prominences as they are called when observed
above the solar limb, can be observed in stable state for many
days or weeks. Sometimes they suddenly start to ascend as a whole
(full eruptions) \citep{Jo11,Ho15}  or
within limited sections of their length (partial eruptions)
\citep{Gi06,Kl14}. The ascending of a
filament can go on high into the corona (successful eruptions) and
gives rise to a coronal mass ejection (CME) or can stop at some
greater height in the corona (confined or failed eruptions) \citep{Ji03,Torok05,Al06,Kur13,Kus15}. Occasionally
two-step eruptions are observed. A filament after the first jump
decelerates and stops at a greater height as in failed eruptions,
but after a rather short period of time it starts to rise again
and develops into the successful eruption with a CME formation.
\citet{By14}  observed on 2011 March 8 at the solar limb the
erupting loop system that stayed in a matastable intermediate
position for an hour and then proceeded and formed the core of a
CME. \citet{Go16}  analyzed observations of the eruption of
a long quiescent filament on 2011 October 22 observed from three
viewpoints by space observatories. A two-ribbon flare and the
onset of a CME appeared 15 hours after the filament disappearance
on the disc. The filament was not observed at the high metastable
position but some coronal structures that can be attributed to a
corresponding flux rope were recognized. A clear example of the
two-step filament eruption on 2015 March 14-15 was reported by
\citet{Wa16}  and \citet{Ch17}. In this event, a
part of a large filament separated from the main body of the
filament at the height of $\approx$  30 Mm and rose upwards to the
height of $\approx$ 80 Mm, where it stayed for 12 hours clearly
visible in chromospheric and coronal spectral lines. Finally it
erupted and produced a halo CME.

Magnetic flux ropes are considered as one of the most probable
magnetic configurations of eruptive prominences. The twisted
structure is often observed in eruptive prominences and cores of
CMEs in a field-of-view of spaceborn coronagraphs \citep{Ga04,Fi08,Jo14,Pa13,Ch14,Gi15}. The magnetic
structure corresponding to a flux rope is measured in space in
magnetic clouds arriving to the Earth's orbit after launches of
CMEs \citep{Le90,Da07}. While some doubts
are raised whether flux ropes exist before eruptions \citep{Ma98,Pa14}, many alternative
configurations transform into flux ropes via reconnection at the
start of the eruptive process \citep{De00,Au10}.

One of the attractive qualities of the flux-rope models is the
possibility of catastrophes in the system equilibrium. \citet{VT78} were first who showed that the equilibrium of a
linear electric current in the coronal magnetic field can be
stable or unstable depending on spatial properties of the coronal
field. \citet{Pr90} analyzed in detail the equilibrium
and dynamics of a straight horizontal flux tube (or line electric
current $I$) in a background magnetic field of a horizontal dipole
$m$ located at the depth $d$ below the conductive surface
(photosphere). Similar model with a vertical dipole was proposed
earlier by \citet{Mo87}. In both cases, there
were found the existence of two equilibrium positions for a rather
small value of the electric current. The lower position was
stable, while the upper position was unstable. When the current
increases, the two equilibrium points approaches closer to each
other. They merge and disappear at the critical height $h_c$ when
the current reaches the critical value. No equilibrium exists in
this field for greater currents. A loss of equilibrium means an
eruption of the filament.

The loss of equilibrium of the straight linear electric current
happens at the point where so called 'decay index' of the coronal
field defined as \citep{Fi00,Fi01} 

\begin{equation}
n = - \frac{\partial \ln B_t}{\partial \ln h},
\end{equation}
where $B_t$ is the tangential to the photosphere component of the
coronal magnetic field and $h$ is the height above the
photosphere, reaches the critical value $n_c = 1$. The loss of
equilibrium was studied in many works \citep{Fo91,Is93,Fo95,Lin98,Lin02,Sc13,Lo14}.

When a twisted flux tube is curved, an extra force is present,
called the hoop force \citep{Sh66,Ba78}. A ring
current is unstable against expansion if the external field
decreases sufficiently rapidly in the direction of the major torus
radius $R$. \citet{Kl06}  following \citet{Ba78} showed,
that it occurs when the background magnetic field decreases along
the expanding major radius of the flux rope $R$ faster than
$R^{-1.5}$. They called the related instability as 'torus
instability'. Thus $n_c = 1.5$ for a thin circular current channel
if the centre is in the photosphere plane. \citet{De10} showed that the critical decay index $n_c$ has similar
values for both the circular and straight current channels in the
range 1.1 - 1.3, if a current channel expands during an eruption,
and in the range 1.2 - 1.5, if a current channel would not expand.
Comparison of the measured heights of stable and eruptive
filaments with the critical heights corresponding to $n_c = 1$,
calculated on the basis of photospheric magnetograms using a
potential magnetic field approximation, showed that the heights of
stable filaments are usually well below the critical heights,
while the heights of filaments just before their eruption are
close to the instability threshold \citep{Fi00,Fi01,Fi08,Fi13,Fi14}.

Two catastrophes occur in the MHD model for the formation and
eruption of solar quiescent prominences proposed by \citet{Zh13}.
The first catastrophe leads to formation of a suspended flux rope
in a quadrupolar magnetic field after  emergence of the rope from below the
photosphere. After the second catastrophe, the quiescent
prominenceeither falls down onto the solar surface
or erupts as a CME. However, the eruption of the prominence is
possible only in a one-step process in this model.

In this paper, we show the possibility of the two-step eruption in
a simple 2D model of the equilibrium of a straight flux tube in the
coronal field with two characteristic spatial scales. We use the
2D approximation because we analyze the initial stages of filament
eruptions when the length of of the filament is much greater than its height above the
photosphere, and the curvature of the tube is small. In the later stages, when the tube takes the shape of a
loop, the hoop force dominates over the diamagnetism of the
photosphere, and 3D models are necessary.

\begin{figure*}
\includegraphics[width=180mm]{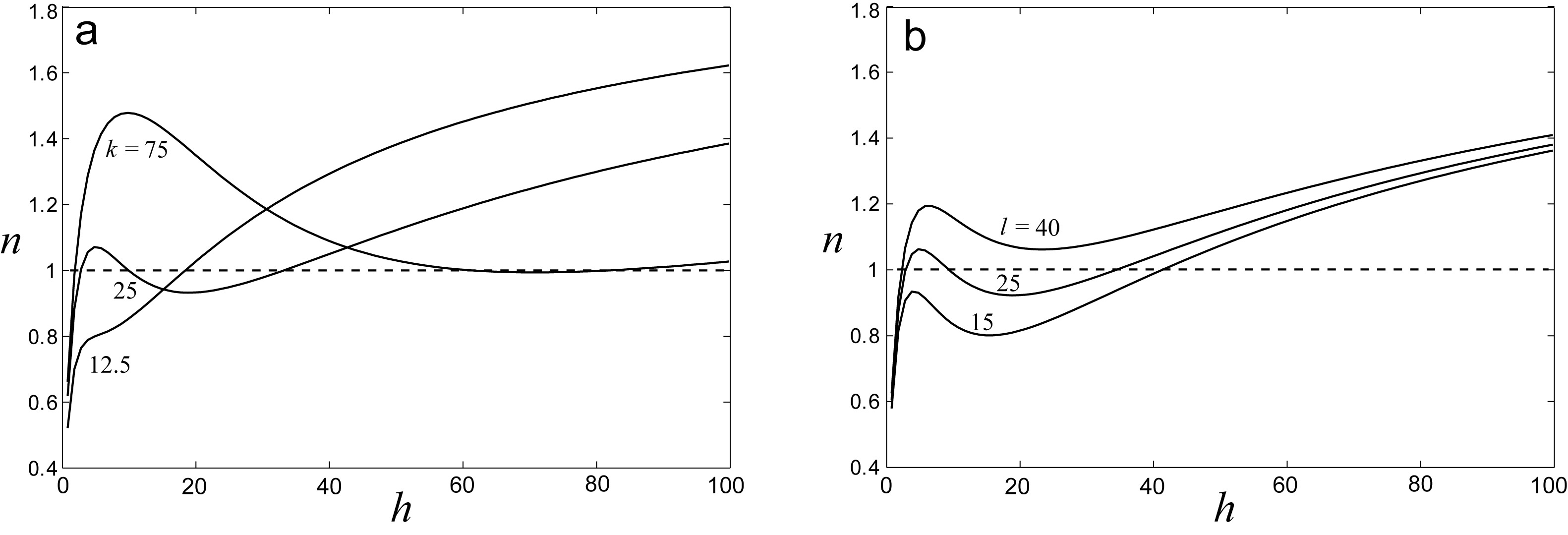}
\caption{ Decay index as a function of height with the value of
the parameter $l = 25$ and different $k$ (a), and with the value
of the parameter $k = 25$ and different $l$ (b). }
\label{fig:n1}
\end{figure*}

\section{Equilibrium of an electric current in a two-scale magnetic field}

We extend the 2D model of \citet{Pr90}   by
introducing another dipole at a larger depth, which produces the
field of a larger scale above the surface. The additional source
is able to create the additional equilibrium point, which may
serve as an intermediate metastable state in a two-step eruption
of a filament.

If the electric current is directed along the $y$-axis direction
at the height $z = h$ within the plane $x = 0$ above the surface
of the photosphere $z = 0$, the magnetic field of the current
outside of the current tube is described by the only one component
of the vector potential ${\bf A} = \{0, A_y, 0\}$

\begin{equation}
A_y^I =  \frac{I}{c} [\log\left(x^2 + (z + h)^2\right)
-\log\left(x^2 + (z - h)^2\right)],
\end{equation}
where the influence of the conductive surface is taken into
account as usual by introducing the mirror current $-I$.

For the equilibrium of an inverse polarity filament in the corona,
the dipolar moments $m_1$ and $m_2$ of both two-dimensional
dipoles should be directed opposite to the $x$-axis. Their vector
potential may be written as

\begin{equation}
A_y^m = - m_1 \frac{z + d_1}{x^2 + (z + d_1)^2} - m_2 \frac{z +
d_2}{x^2 + (z + d_2)^2},
\end{equation}
where $d_1$ and $d_2$ are depths of the dipoles below the
photosphere. Both dipoles are located also in the plane $x = 0$.
Then the horizontal component of the force acting on the current
tube vanishes according the symmetry, while the vertical component
(per unit length) is \citep{VT78,Mo87,Pr90} 

\begin{equation}
F_z(h) =  \frac{I^2}{c^2 h} - B_x^m(h) \frac{I}{c} - M g,
\end{equation}
where $M$ is the mass of the tube per unit length, $g$ is the free
fall acceleration,

\begin{equation}
B_x^m(z) = - \frac{\partial A_y^m}{\partial z}.
\end{equation}
Equilibrium is achieved at any given height $h$ with the value of
the current as

\begin{equation}
I_0 = c \left(\frac{B_x^m h}{2} \pm \sqrt{\frac{{B_x^m}^2 h^2}{4} +
M g h}\right),
\end{equation}
The sign (--) before the square root corresponds to a normal
polarity filament, not interesting for our study. To solve
Equation (4) with the zero left side analytically relative $h$ is
not so easy even neglecting the last gravitational term, since we
have quartic equation. We may expect, however, that there are two
critical heights in this field, each of them corresponding to the
scale of the field of each source $m_1$ and $m_2$.

It was shown \citep{VT78,Fi00,Fi01}  that a constant straight linear current becomes unstable
when the decay index of the background field exceeds unit:

\begin{equation}
n = - \frac{\partial \ln B_x}{\partial \ln h} > 1.
\end{equation}
In the field of a single dipole $m$, the decay index $n$ changes
with height $h$ as

\begin{equation}
n (h) =  \frac{2 h}{h + d}.
\end{equation}

For two dipoles (3) it is equal to

\begin{equation}
n(h^\prime) = 2\frac{\left(\frac{h^\prime +k}{h^\prime +1}
\right)^3 +l}  {\left(\frac{h^\prime +k}{h^\prime +1} \right)^2 +
l} \frac{h^\prime}{h^\prime + k},
\end{equation}
where $h^\prime = h/d_1$, $k = d_2/d_1$, $l = m_2/m_1$. Since it
is difficult to solve the equation

\begin{equation}
n(h^\prime) = 1
\end{equation}
analytically, we will analyze the right part of Equation (9)
numerically in order to choose parameters $k$ and $l$ providing
multiple solution of Equation (10).

Figure 1 shows the dependence of the decay index described by
Equation (9) on height with different values of the parameters $k$
and $l$. When $k$ is relatively small, the dependence is monotonic,
and Equation (10) has only unique solution (Fig. 1a). When $k$ is
too large, the curve has extremums but it crosses the line $n = 1$
only one time at a low height. Thus we choose $k = 25$, which
provides three distinct roots of Equation (10). Similarly,  we choose the value of the parameter $l = 25$ (Fig. 1b).
Then we set all dimensionless parameters of the model as $m_1 = -1,
m_2 = -25, d_1 = 2, d_2 = 50, M = 10^{-7}, g = 30$. Our intention is to
describe the equilibrium of a flux rope in the solar corona at a
height of about 20 Mm in a magnetic field of about 10 gauss with the
gravity force of about 1\% of the Lorenz force. This gives that
the dimensionless units correspond to: length - 10 Mm, time -
$10^3$ s, mass - $10^{12}$ g, magnetic induction - 1 gauss. Figure 2
shows the field lines of the model with the equilibrium electric
current (6) at the height of 1.5. The field lines are represented
by isocontoures of the condition $A_y^I + A_y^m = const$.

Figure 3 shows the value of the equilibrium electric current $I_0$ as
a function of height according to Equation (6). Crossings of the
curve with horizontal lines determine equilibrium heights. There
are intervals of current values where four, two or no equilibrium
positions exist. Crossings with rising parts of the curve
correspond to the stable equilibrium; crossings with descending
parts correspond to the unstable equilibrium. Two maxima of the
curve, at $ h \approx 3$ and $h \approx 35$, indicate the places
where catastrophic losses of equilibrium can happen. They are two
critical heights in the model. As it is seen in Fig. 3, the value
of the critical current at the lower point is a little smaller
than the critical current at the higher point.

\begin{figure}
\includegraphics[width=80mm]{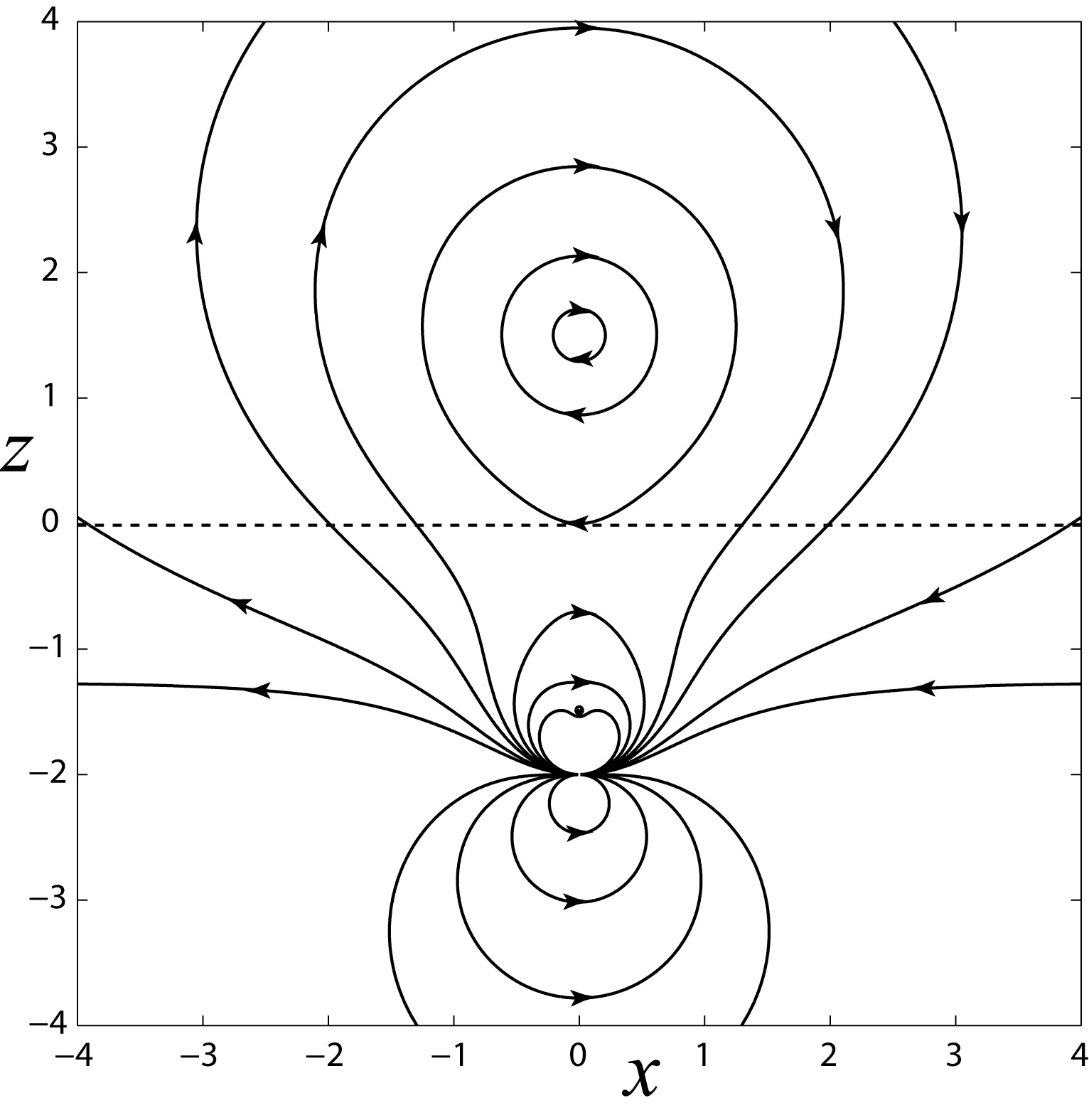}
\caption{Magnetic field lines of the model with the equilibrium
electric current at the height of 1.5.}
\label{fig:FL}
\end{figure}

\begin{figure}
\includegraphics[width=80mm]{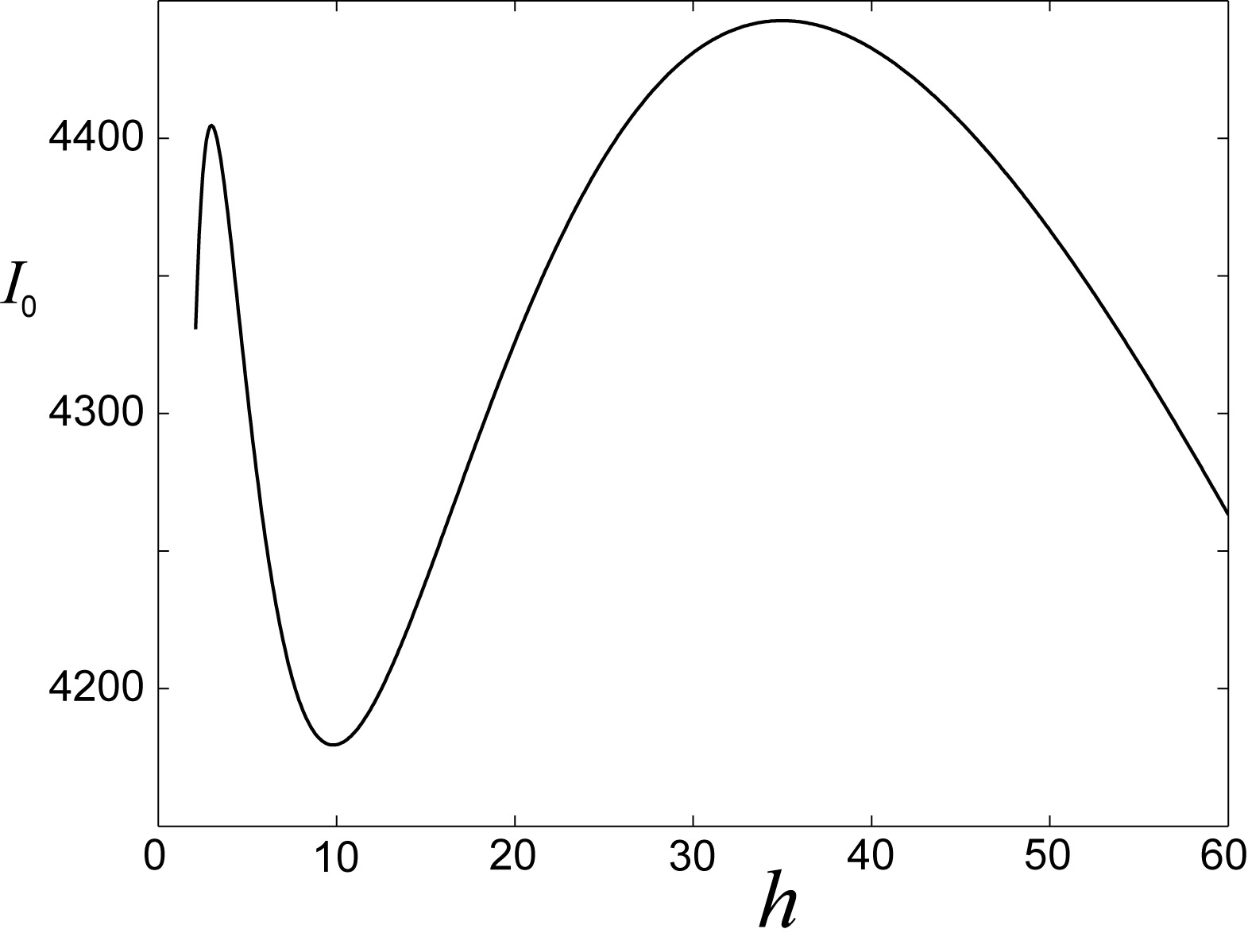}
\caption{Value of the equilibrium electric current $I_0$ as a
function of height according to Equation (6). } \label{fig:I0}
\end{figure}

\section{Dynamics of an electric current in the two-scale magnetic field}

Let us consider the equation of motion of the current tube in the
two-scale magnetic field:

\begin{equation}
M \frac{d^2 h}{dt^2} =  F_z(h) - \eta \frac{d h}{d t},
\end{equation}
where $\eta$ is the coefficient of artificial dissipation
(viscosity) introduced in order to avoid long oscillations of the
tube. We solve Equation (11) numerically using MATLAB solver
'ode45' based on an explicit Runge-Kutta  formula. Figure 4 shows
the relaxation of the tube after small disturbance near the
equilibrium position at $h = 1.5$. The value of $\eta$ is chosen
as $10^{-6}$, which means that the decelerating force at the speed
of 1000 km s$^{-1}$ is about 1\% of the Lorenz force.

The presence of the additional dipole leads to the change of the
critical height for the instability from $h_c = 2$ to $h_c \approx
3$. We then put the current tube a little bit lower the critical
height and change the both dipolar moments as

 \begin{equation}
m_i = {m_0}_i \exp\left(- \frac{t^2}{\tau^2}\right),
\end{equation}
with $\tau = 10^3$ in order to analyze the eruption of the tube.
For finite displacements of the current tube we should take into
account the changes of the current due to inductance.  \citet{Pr90}  considered several possibilities to determine
current $I$ as a function of $h$ (or $t$).

\begin{figure*}
\includegraphics[width=140mm]{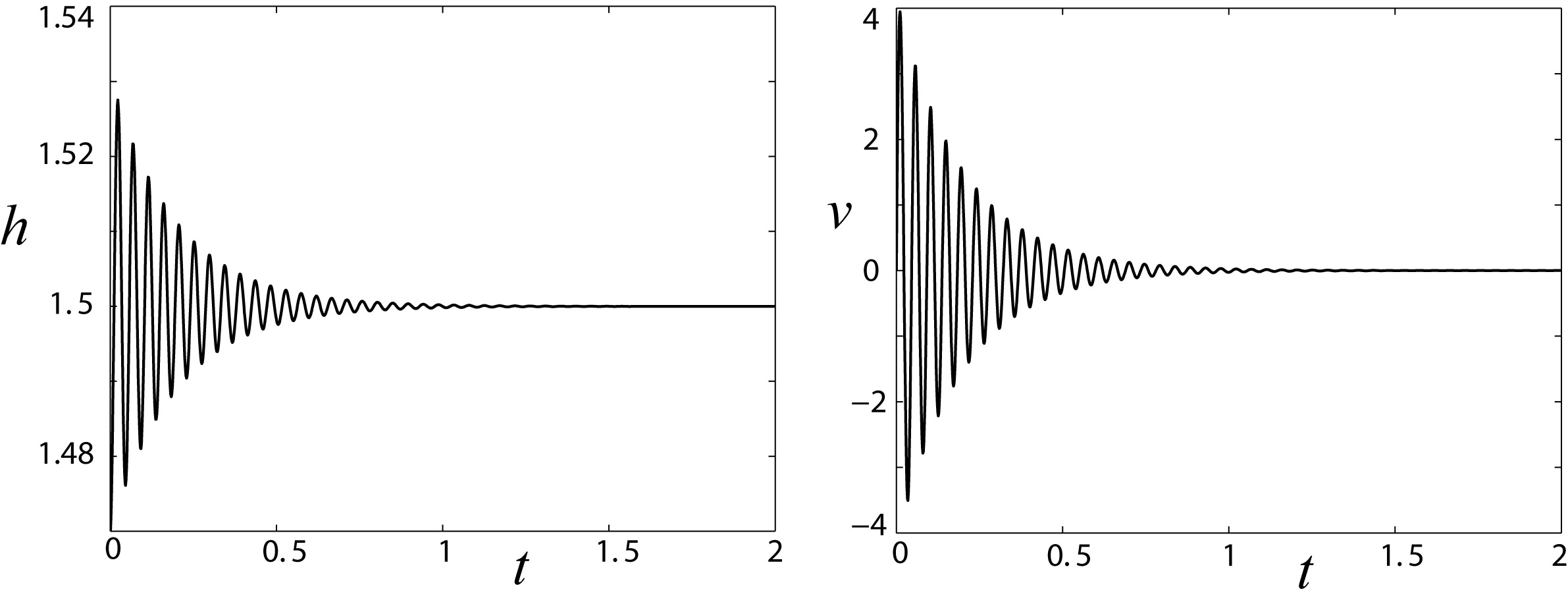}
\caption{ Relaxation of the tube after small disturbance near the
equilibrium position at $h = 1.5$. } \label{fig:S}
\end{figure*}

\subsection{Constant current}

The simplest assumption is $I = const$. \citet{Pr90} 
decided this case not the most realistic because it needs the
source of energy to support the constancy of the current. However,
in a case of a partial eruption, when only some section of the
long flux rope is moving, the large self-inductance of the circuit
can easily support the current to be nearly constant.

\begin{figure*}
\includegraphics[width=180mm]{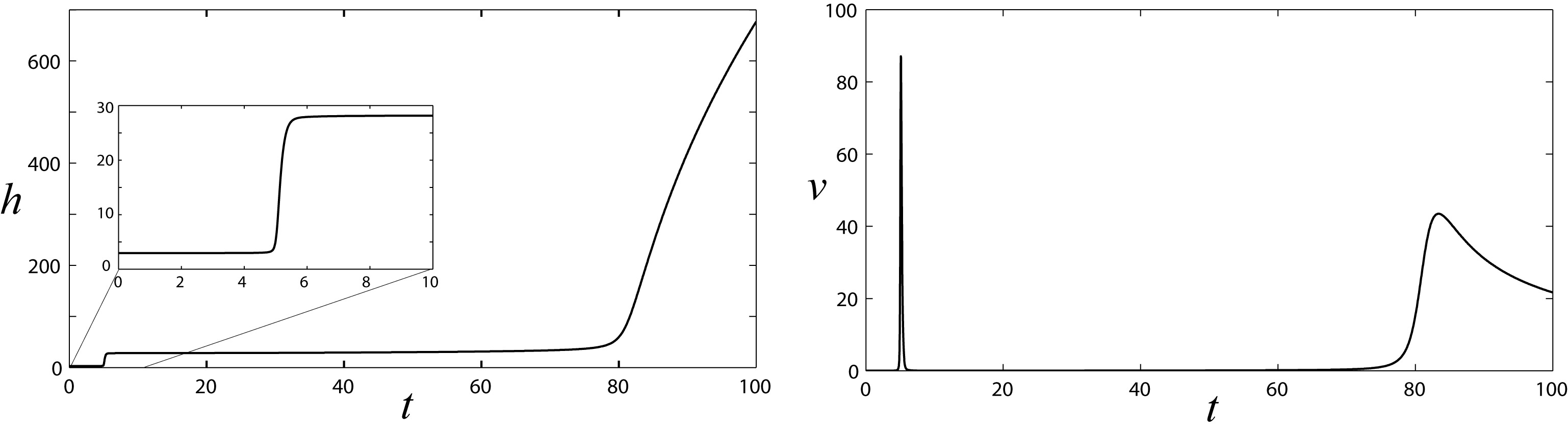}
\caption{ Height-time (left) and velocity-time (right) plots of
the dynamics of the tube with the constant current.}
\label{fig:I0_const}
\end{figure*}

Figure 5 shows the dynamics of the tube with the constant current
starting from the point $h_0 = 2.95$ close to the critical height.
After slow rising due to the decreasing the background field, the
tube erupts very rapidly at $t = 5$ but then decelerates and stops
at the height of $h = 28$. The tube stays approximately at this
height, slowly rising again till to $t = 80$. Then the next
eruption happens.

The critical height in the field of a single dipole is $h_c = d$,
while the critical value of the current is

\begin{equation}
I_c =\frac{c m}{4 d}.
\end{equation}

The presence of another dipole raises the critical height in the
smaller-scale field from $h_c = 2$ to $h_c \approx 3$ and lowers
the critical height in the greater-scale field from $h_c = 50$ to
$h_c \approx 30$. Since we have chosen in our model $m_1/d_1 =
m_2/d_2$, the values of the critical current according Equation
(13) are equal in both separate fields. In the total field, the
critical current value at the higher critical
height is a little higher than the
critical current  at the lower critical
height. That is why the tube is able to stop at the intermediate
height of  $h \approx 30$ and to start the next eruption after the
decrease of the background field by about 1\%.

\subsection{Constant magnetic twist}

If we consider the cylindrical tube as a part of a rising
three-dimensional loop with the ends anchored in the photosphere, a
reasonable approximation is the constant net twist along the tube.
\citet{Pr90} showed that this condition leads to the
inversion dependence of the current on height

\begin{equation}
I = I_0\frac{h_0}{h},
\end{equation}
where $I_0$ is the initial current and $h_0$ is the initial height.
Eruption becomes impossible under this condition because the
equilibrium is stable at any height because the repulsive force
(the first term in the right part of Equation (4)) decreases with
height faster ($\sim h^{-3}$) than the attractive force (the
second term, $\sim h^{-1}(h + d)^{-2}$) even neglecting the
gravity.

\subsection{Constant magnetic flux}

A reasonable condition for determining $I$ is to fix the magnetic flux between the
photosphere and the flux tube. Thanks to the translational
symmetry the flux is determined by the vector potential

\begin{equation}
\Phi = A_y(0) - A_y(h - r_0),
\end{equation}
where $r_0$ is the radius of the flux tube. Using Equations
(2)-(3) and taking into account that $r_0 \ll h$ we have

 \begin{equation}
\Phi = -\frac{2 I}{c} \log \frac{2 h}{r_0} - m_1 \frac{h}{d_1(d_1
+ h)} - m_2 \frac{h}{d_2(d_2 + h)} .
\end{equation}

The condition $\Phi = const$ determines the dependence of the
current value $I$ on height from Equation (16), if we specify the
radius of the flux tube $r_0$. Figure 5 shows different vertical
profiles of the current under various conditions. The lowest curve
corresponds to the constant magnetic twist. Two other descending
curves show changes of the current in the tube with a constant
radius $r_0 = 0.01 h_0$ and $r_0 = 0.001 h_0$.  Two upper curves
correspond to the tubes with the same initial radii but linearly
expanding with height. The smaller the tube radius, the weaker the
dependence of the current with  height.

\begin{figure}
\includegraphics[width=\columnwidth]{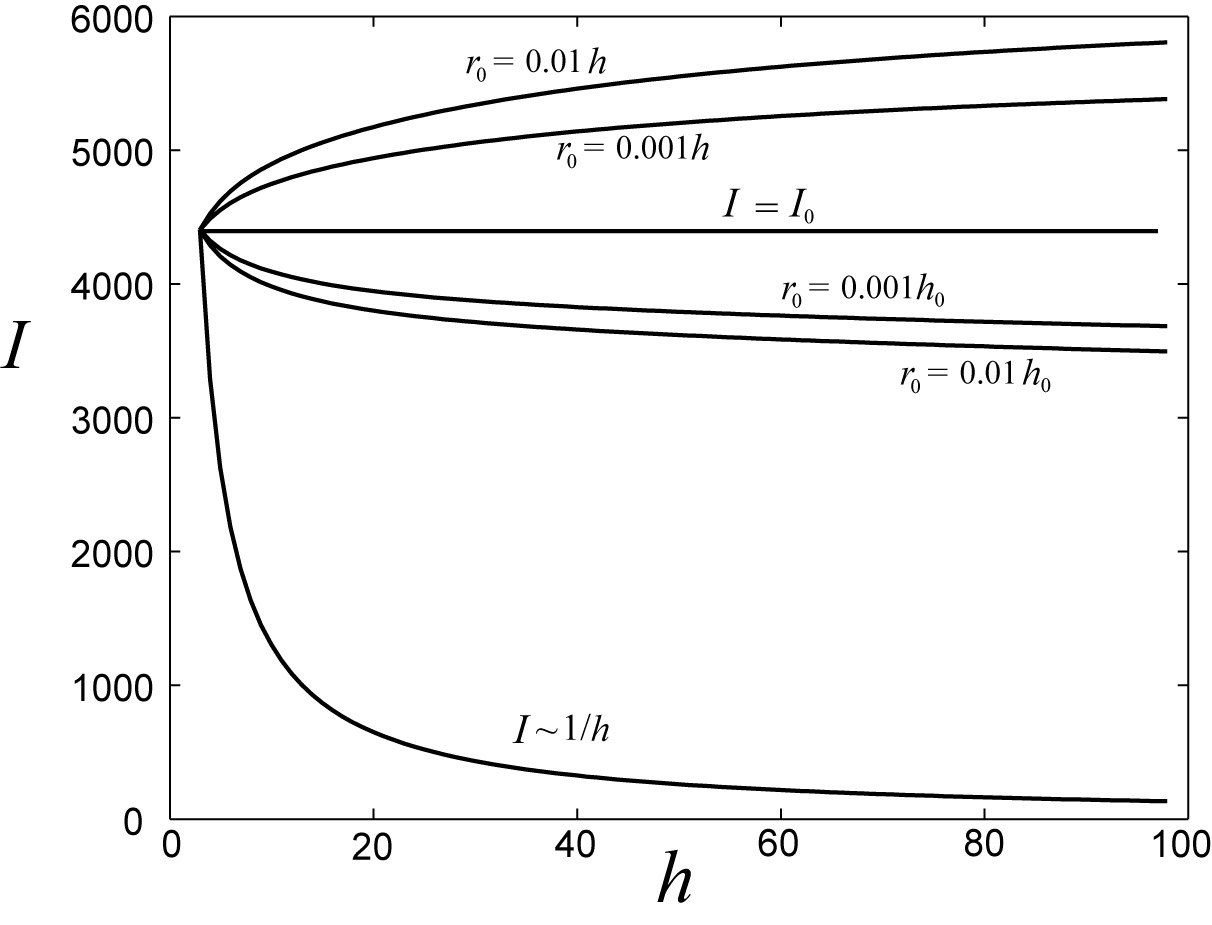}
\caption{ Electric current as a function of height under different
conditions.}
\label{fig:I(h)}
\end{figure}

\subsubsection{Constant magnetic flux-tube radius}

The condition of the constant magnetic flux between the
photosphere and the magnetic flux tube with a constant radius leads
to decreasing of the current with height (Fig. 5). Figure 7 shows
the rise of the tube of the constant radius $r_0 = 0.001 h_0$ from the
equilibrium position $h_0 = 2.95$, while Figure 8 corresponds to
the radius $r_0 = 0.01 h_0$. In both cases, eruptions start at
later time and a larger height than in Fig. 5 since the current is
decreasing with height. We see the two-step eruption similar to
Fig. 5, but the jump of the tube in the first step is smaller
(from $h \approx 3$ to $h \approx 8$) in Fig. 7 and very smooth in Fig.
8. The second-step eruption starts at $t \approx 450$ and $t
\approx 550$, respectively. Velocities are lower than in Fig. 5 in
the whole event, with the first peak lower than the second one in
contrast to Fig. 5. Especially slow is the first eruption in Fig.
8.

\begin{figure*}
\includegraphics[width=180mm]{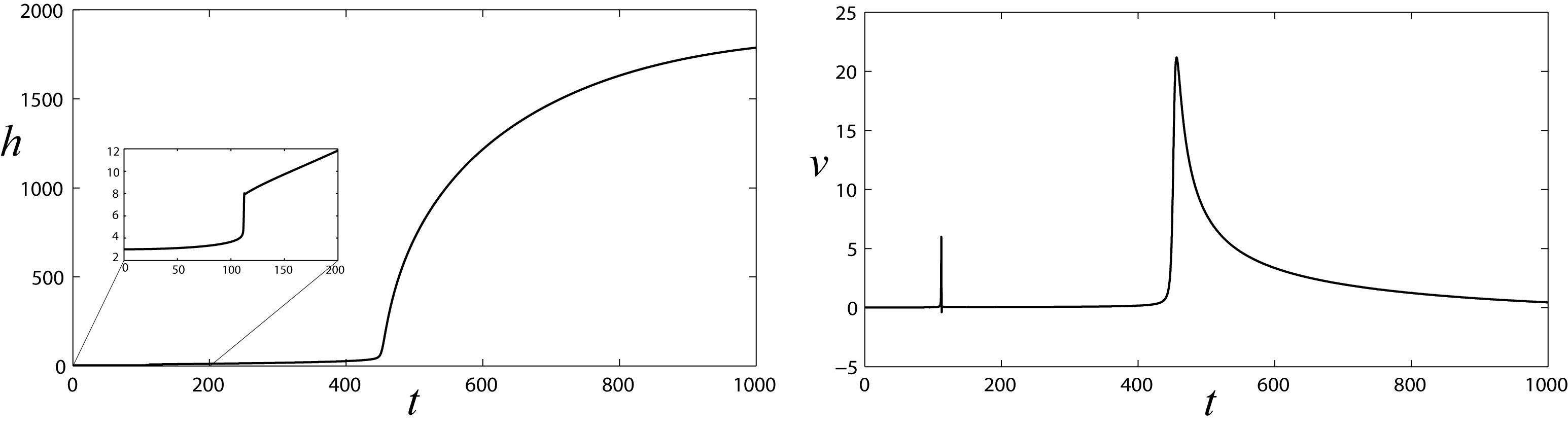}
\caption{Height-time (left) and velocity-time (right) plots of the
dynamics of the tube with the constant radius $r_0 = 0.001 h_0$.}
\label{fig:r0_001}
\end{figure*}

\begin{figure*}
\includegraphics[width=180mm]{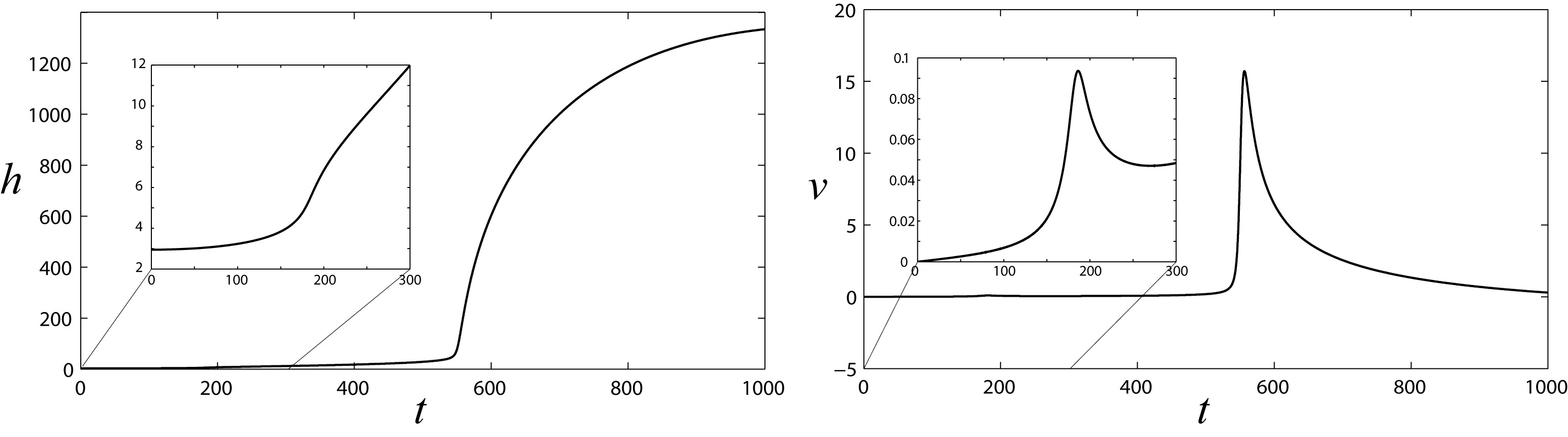}
\caption{The same as in Figure 7 with the radius $r_0 = 0.01
h_0$.} \label {fig:r0_01}
\end{figure*}

\subsubsection{Expanding flux-tube}

If the radius of the tube expands linearly with height, the
current slowly increases with height (Fig. 6). Figure 9 shows the
dynamics of the tube with $r_0 = 0.01 h$ and $r_0 = 0.001 h$  from
the equilibrium position $h_0 = 2.95$. The eruption starts
violently at $t \approx 0.5$ and does not show the two-step
scenario. The tube obtains too high velocity, and the current
increases too much for obtaining a higher equilibrium position. It
is possible to put the tube at the higher equilibrium point on the
rising section of the equilibrium curve in Fig. 3, but the
eruption in this case is also a one-step eruption in the
larger-scale field.

\begin{figure*}
\includegraphics[width=120mm]{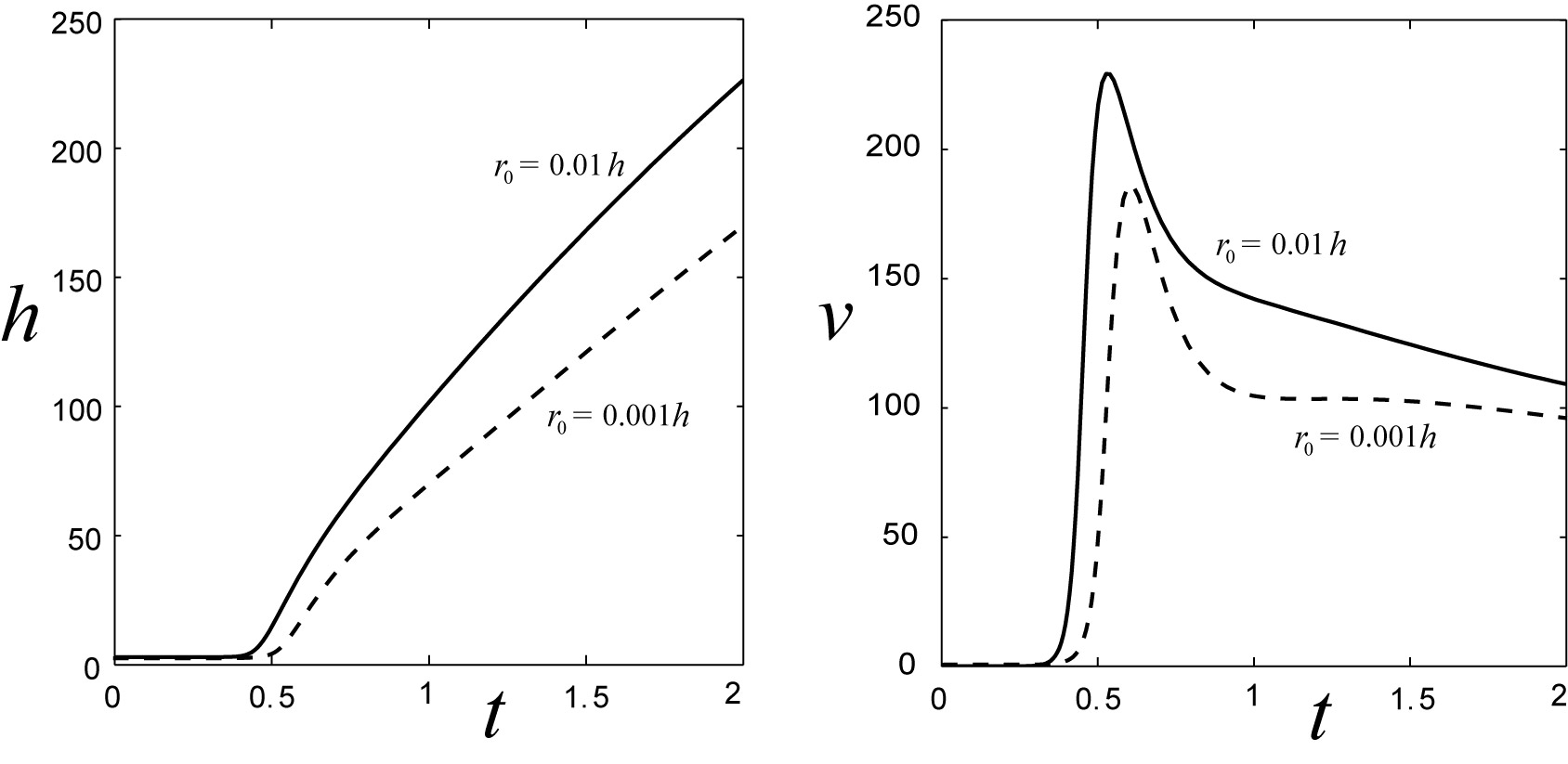}
\caption{Height-time (left) and velocity-time (right) plots of the
dynamics of the tube with the linearly increasing radius $r_0 =
0.001 h$ (solid lines) and $r_0 = 0.01 h$ (dashed lines).}
\label{fig:r0_h}
\end{figure*}

\section{Discussion and Conclusions}

We analyzed a possibility of two-step eruptions using a simple 2D
model of a twisted flux tube equilibrium in the two-scale coronal
magnetic field. This field is created in the model by two 2D
dipoles at different depths below the photosphere. We chose
parameters of the dipoles in order to have four equilibrium
positions for some values of the electric current within the flux
tube. Two of them are stable and the other two are unstable. There
are two values of the current when a sudden loss of equilibrium
happens. Each critical value of the current corresponds to the
critical height of the current above the photosphere. We tried to
choose the parameters so that the critical heights are quite
different but the critical current values are similar. Then, if
the coronal field slowly decreases, the tube erupts starting from
the lowest critical height. The following evolution of the tube
depends on changes of the electric current within it. Owing to
induction under different assumptions, it may increase, decrease
or to be constant (Fig. 6). If some dissipation is present in the
system, the tube can stop at the height between the two critical
heights. After rather short time spent in the metastable state,
the tube can erupt again from the higher critical height. Thus,
the model shows possibility of the two-step eruption under
reasonable conditions.

We can compare the vertical profile of the decay index in the
model (Fig. 1) with the real distribution of the decay index in
the region that produced the two-step eruption \citep{Go16} (Gosain et al.
2016). Figure 10 shows the distribution of the decay index $n$
calculated in a potential-field approximation on the basis of {\it
SDO}/HMI magnetogram (the Helioseismic and Mangetic Imager (HMI) \citep{Sc12}
onboard {\it Solar Dynamics Observatory
(SDO)}) on 2011 October 18 at 00 UT in the horizontal plane at the
height of 70 Mm and along the vertical line above the pre-eruptive
position of the filament. The vertical profile has the similar
shape as shown in Fig. 1. There are two zones of stability $(n <
1)$: one at low heights $h \lesssim 100$ Mm and the second at
heights between 350 and 500 Mm.

\begin{figure*}
\includegraphics[width=140mm]{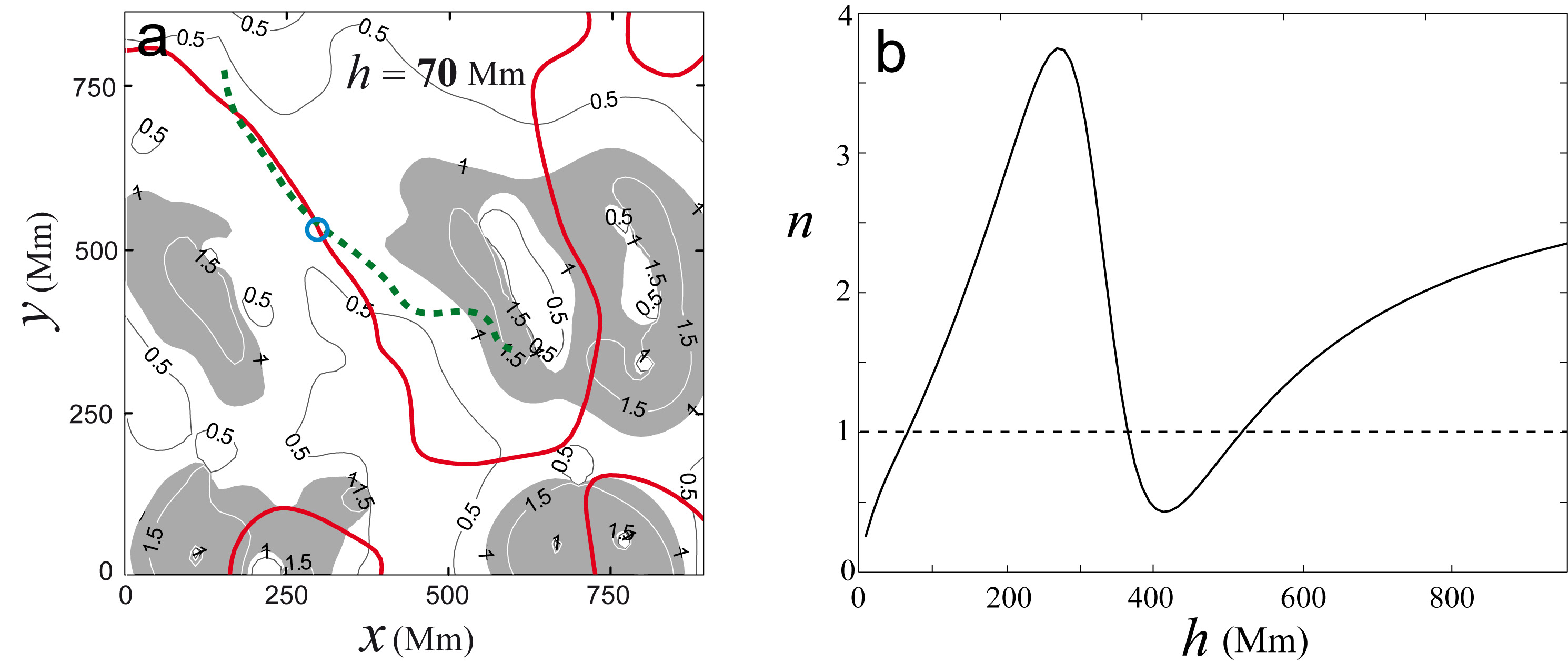}
\caption{Distribution of the decay index $n$ in the $xy$-plane at
the height of 70 Mm (a) and along the vertical line above the blue
circle shown in the left panel (b) calculated in a potential-field
approximation on the basis of {\it SDO}/HMI magnetogram on 2011
October 18 at 00 UT. The dashed green line indicates the position
of the quiescent filament.} \label{fig:n_ex}
\end{figure*}

The eruption of the quiescent filament, shown by the dashed green
line in Fig. 10a, started at about 03:40 UT on 2011 October 21. The
initial height of the filament was about 100 Mm, close to the
first cross of the curve with the line $n = 1$ in Fig. 10b. The
filament stopped at the height of about 350 Mm \citep{Go16},
which corresponded to the stability zone $(n < 1)$ in Fig. 10b.
While most of the filament material drained back to the
chromosphere shortly after it had stopped, the faint coronal
structure was already recognizable before the filament started to ascend again at about 21 UT, more than 17 hours after the start of
the first step. The second-step eruption might start near the
upper border of the zone of stability at the height of about 500
Mm (Fig. 10b). Thus, the scenario of the event on 2011 October 21
is very similar to the predictions of our model. The two-step
eruption owes to the special distribution of the photospheric
magnetic field in the region, which manifests itself in
owes its existence to the special distribution of the photospheric magnetic field in the region, which manifests itself in the non-monotonic behavior of the decay index in the corona.

An understanding of mechanisms of the two-step eruptions is significant for the perception of the whole picture of solar eruptive events and space weather implications. The large delay of a CME after the filament eruption start may influence on correlation statistics between the two manifestations of an eruption. For the on-disc events that result in faint halo CMEs, estimations of a CME arrival to the Earth on the basis of observations of filament activations may be not very precise. The structure of the coronal magnetic field influences essentially both on the time profile of the filament ascending and its trajectory in the corona.

\section*{Acknowledgements}

The author is very grateful to the reviewer Prof. T. Forbes for giving his valuable suggestions for improving the manuscript. The author thanks the {\it SDO}/HMI team for the high-quality data
supplied.




\bibliographystyle{mnras}
\bibliography{reference} 



\bsp    
\label{lastpage}
\end{document}